\documentclass[prl,twocolumn,superscriptaddress]{revtex4-2} 
\usepackage{mathrsfs}
\usepackage{amsfonts}
\usepackage{amsmath}
\usepackage{txfonts}
\usepackage{amssymb}
\usepackage{graphicx,subfigure}
\usepackage{bm}
\usepackage{color}
\usepackage[normalem]{ulem}
\usepackage{dcolumn} 
\usepackage{bbold}
\usepackage[table]{xcolor}
\usepackage{float}
\usepackage{tabularx}

\definecolor{pacificb}{HTML}{1CA9C9}

\begin{document}

\title{
Harnessing the skyrmion Hall effect for low-power skyrmion 
transport in a tubular synthetic antiferromagnet
}

\author{Ivan P. Miranda}
\affiliation{Department of Mathematics and Physics, Linnaeus University, Kalmar, Sweden}
\email{ivan.depaulamiranda@lnu.se}

\author{Grzegorz J. Kwiatkowski}
\affiliation{Science Institute, University of Iceland, 107 Reykjav\'ik, Iceland}

\author{Jacob J. Mankenberg}
\affiliation{Department of Mathematics and Physics, Linnaeus University, Kalmar, Sweden}

\author{Cecilia M. Holmqvist}
\affiliation{Department of Mathematics and Physics, Linnaeus University, Kalmar, Sweden}

\author{Igor S. Lobanov}
\affiliation{Department of Physics, ITMO University, 197101 St. Petersburg, Russia}

\author{Valery M. Uzdin}
\affiliation{Department of Physics, ITMO University, 197101 St. Petersburg, Russia}

\author{Pavel F. Bessarab}
\affiliation{Department of Mathematics and Physics, Linnaeus University, Kalmar, Sweden}
\affiliation{Science Institute, University of Iceland, 107 Reykjav\'ik, Iceland}
\email{pavel.bessarab@lnu.se}

\date{\today}

\begin{abstract}
We show that optimal control can turn the skyrmion Hall effect from a source of typically unwanted transverse motion into a mechanism for reducing Ohmic losses in skyrmion transport. We consider a pair of antiferromagnetically coupled skyrmions confined to a tubular geometry, where their relative transverse displacement becomes a periodic internal coordinate. The skyrmion Hall effect drives this coordinate, while finite interlayer coupling makes the resulting relation between applied current and longitudinal velocity nonlinear, allowing different current protocols to produce the same prescribed average velocity. We determine which of these protocols minimizes Joule heating. Depending on the interlayer coupling and spin-transfer-torque parameters, the optimal current is either constant, with the skyrmion pair maintaining a fixed relative position, or time dependent, with the pair undergoing periodic relative motion around the tube. In either case, the additional internal degree of freedom created by the skyrmion Hall effect and interlayer coupling enables skyrmion transport at a lower power than in the uncoupled limit.
\end{abstract}

\maketitle

\textit{Introduction.---}Skyrmions -- topologically protected spin textures that naturally emerge in chiral magnets and in systems with sizable Dzyaloshinskii-Moriya interaction (DMI) -- have been considered promising information carriers for low-power spintronic devices over the last decade \cite{Fert2017,Luo2018,Bogdanov2020}. A central challenge to their technological deployment is the control of their motion, required by many potential applications, such as logic devices \cite{Zhang2015}, racetrack and shift-register memories, and hybrid skyrmionic-electronic devices \cite{Song2020}. One way to achieve this control is to use spin-polarized electric currents via the spin-transfer torque (STT) effect \cite{Nagaosa2013,Fert2013}. In fact, an important characteristic of skyrmions is that they can be moved by lower currents than other spin textures, such as domain walls \cite{Muhlbauer2009}, thereby reducing the energy cost associated with Joule heating.

Despite this notable property, skyrmions in single layers also face an undesired effect. Current-induced torques deflect them from their intended path -- a phenomenon known as the skyrmion Hall effect (SkHE) \cite{Chen2017} -- causing edge losses at technologically relevant speeds. The SkHE is a direct consequence of the finite topological charge of skyrmions. Since its theoretical description and experimental detection, several strategies have been proposed to overcome this obstacle \cite{Yang2024}. These include antiferromagnetic (AFM) skyrmions \cite{Barker2016} in synthetic antiferromagnets (SAFs) \cite{Zhang2016, Clecio2025, Pham2024, Dohi2019, Legrand2019}, which exploit zero net topological charge, and single skyrmions in tubular racetracks \cite{Wang2019}, which are edgeless by geometry and increasingly viable due to advances in 3D nanofabrication techniques \cite{Sheka2021,Makarov2021,Farinha2025}. In terms of tunability, the advantage of such solutions is natural: both SAFs and tubular racetracks emerge as versatile platforms with a higher degree of controllability in comparison with other solutions to the SkHE problem, such as bulk antiferromagnets \cite{Gao2020,Jani2021} or skyrmioniums on single planar racetracks \cite{Zhang2016skyrmionium, Kolesnikov2018}. In SAFs, for instance, the interlayer exchange, with Ruderman-Kittel-Kasuya-Yosida (RKKY) \cite{Ruderman1954, Kasuya1956, Yosida1957} character, can be controlled by the spacer thickness. In tubes, system parameters such as radius and thickness \cite{Wang2019} also drive key dynamical properties, such as the skyrmion angular speed. However, from the control side, the response to the current remains trivial since the skyrmion drift velocity is simply a linear function of the applied current \cite{Barker2016,Wang2019}. In other words, in these settings, the current does not activate an internal transport channel beyond uniform drift.


Motivated by these appealing properties, but also by this remaining limitation, a natural question arises: \textit{Can the current excite an additional useful mode on such topological textures in order to reduce Joule heating losses even further?} Since SAFs consist of a pair of oppositely charged skyrmions, a transverse separation mode can be intrinsically excited by the opposing SkHEs acting on the two skyrmions. In a planar geometry, this mode can drive the pair toward racetrack edge annihilation and compromise the main advantage of AFM skyrmion pairs in SAFs (SAF-SPs); for this reason, this mode, often referred to as skyrmion pair \textit{decoupling}, has been regarded as an undesired feature of this setup \cite{Zhang2016}. But what if we could combine SAFs with the tubular geometry solution?

Here, 
we show that SAF-SPs in a tubular geometry constitute a qualitatively distinct system. In this case, the separation mode becomes a controllable periodic degree of freedom and the drift velocity becomes a nonlinear function of the applied current, with skyrmion deformation providing an important source of nonlinearity. To demonstrate this, we describe the dynamics using a two-coordinate collective model obtained by projecting the Landau-Lifshitz (LL) equation with Zhang-Li (ZL) \cite{Zhang2004} STT onto the slow longitudinal and transverse modes. By applying optimal control theory, we show that the separation mode changes the structure of the minimal dissipation problem itself: 
the minimum-power solution within the \textit{locked} regime, where the SAF-SP is coupled at a fixed distance between the centers and driven by dc currents, lowers the typical power scale compared to the uncoupled case by at least a factor $\alpha^{2}$ -- where $\alpha$ is the Gilbert damping constant. Concurrently, the \textit{winding} regime, where the distance between the SAF-SP centers varies periodically, opens finite regions in the STT parameter space in which time-dependent currents can outperform the locked dc solution. Thus, in a tube, the SkHE is effectively converted from an internal instability into a functional low dissipation transport channel.

\begin{figure}[t!]
    \centering
    \includegraphics[width=1\linewidth]{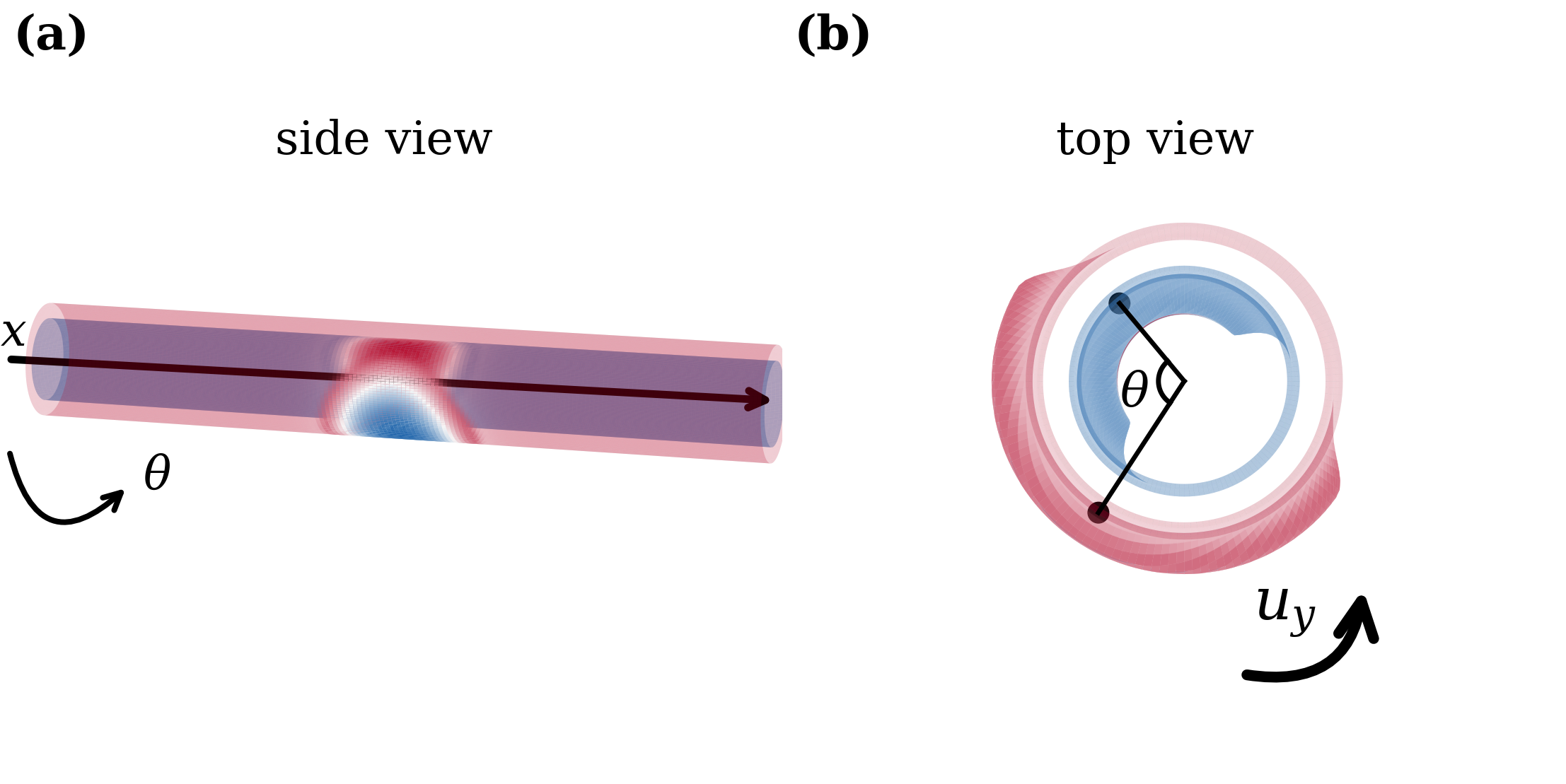}
    \caption{\textbf{Sketch of the SAF-SP on a tube.} (a) side view; (b) top/front view. The separation coordinate $u_y$ can be conveniently mapped onto the angular coordinate $\theta$. The current is applied along the longitudinal direction, $x$: $\vec{j}=j_x\vec{e}_x$.}
    \label{fig:sketch-system}
\end{figure}

\textit{Model.---} We consider two identical ferromagnetic racetracks antiferromagnetically coupled, each hosting a single metastable skyrmion, as sketched in Fig.~\ref{fig:sketch-system}. The system consists of a lattice of spins with spacing $a=1$\AA$\,$ and tubular geometry, simulated here by imposing periodic boundary conditions in the transverse direction. The spin texture in each layer is described by an atomistic spin Hamiltonian $\mathcal{H}$ containing intralayer exchange ($\mathcal{J}_{\parallel}$), 
interlayer exchange ($\mathcal{J}_{\perp}$), Dzyaloshinskii-Moriya interaction (DMI, $\vec{\mathscr{D}_{ij}}$), and uniaxial anisotropy ($\mathcal{K}_{\parallel}$):
\begin{equation}
\label{eq:hamiltonian}
\begin{gathered}
\mathcal{H} = -\mathcal{J}_{\parallel} \sum_{\substack{\ell i\\ j \in \mathscr{N}(i)}}
\vec{e}_{\ell i}\cdot\vec{e}_{\ell j}
-\mathcal{J}_{\perp}\sum_i
\vec{e}_{1i}\cdot\vec{e}_{2i}\\
-\sum_{\substack{\ell i\\j \in \mathscr{N}(i)}}
\vec{\mathscr{D}}_{ij}\cdot
\left(\vec{e}_{\ell i}\times\vec{e}_{\ell j}\right)
+\mathcal{K}_{\parallel}\sum_{\ell i}
\left(\vec{e}_{\ell i}\cdot\vec{e}_z\right)^2 .
\end{gathered}
\end{equation}

\noindent Here, $i$ and $j$ are site indices, $\ell=\{1,2\}$ is the layer index, and $\vec{e}_{\ell i}$ denotes the magnetic moment direction of site $i$ in layer $\ell$. The exchange constants are $\mathcal{J}_{\parallel}>0$ and $\mathcal{J}_{\perp}<0$, with $|\mathcal{J}_{\perp}|$ varied in the range $[0, 6\times 10^{-3}\mathcal{J}_{\parallel}]$. The DMI vector $\vec{\mathscr{D}}_{ij}$ lies in the layer plane and is perpendicular to the bond direction $\vec{r}_{ij}=\vec{r}_i-\vec{r}_j$, all with the same magnitude $|\vec{\mathscr{D}}_{ij}|=\mathscr{D}_{\parallel}=0.12\mathcal{J}_{\parallel}$. The anisotropy constant is $\mathcal{K}_{\parallel}=-0.05\mathcal{J}_{\parallel}$, and $\mathscr{N}(i)$ denotes the set of nearest neighbors of site $i$. The dynamics of $\vec{e}_{\ell i}$ is then described by the LL equation including ZL STT \cite{Zhang2004},
\begin{equation}
\label{eq:ll-torque}
\begin{gathered}
\dot{\vec{e}}_{\ell i}=-\gamma'\vec{e}_{\ell i}\times\vec{B}_{\ell i}
-\alpha\gamma'\vec{e}_{\ell i}\times
\left(\vec{e}_{\ell i}\times\vec{B}_{\ell i}\right)\\
-\frac{\alpha\zeta_1-\zeta_2}{1+\alpha^2}
\vec{e}_{\ell i}\times(\vec{j}\cdot\nabla)\vec{e}_{\ell i}
+\frac{\zeta_1+\alpha\zeta_2}{1+\alpha^2}
\vec{e}_{\ell i}\times
\left[\vec{e}_{\ell i}\times(\vec{j}\cdot\nabla)\vec{e}_{\ell i}\right],
\end{gathered}
\end{equation}
where $\gamma'=\gamma/(1+\alpha^2)$ is the rescaled gyromagnetic ratio, $\vec{B}_{\ell i}=-\mu^{-1}\partial\mathcal{H}/\partial\vec{e}_{\ell i}$ is the effective magnetic field with $\mu=1$ $\mu_B$ being the magnitude of the magnetic moment, $\vec{j}$ is the applied current density, and $\zeta_{1,2}$ are material-specific STT parameters from which the nonadiabaticity parameter $\beta=\zeta_2/\zeta_1$ is defined. We take the current along the longitudinal direction, $\vec{j}=j_x\vec{e}_x$, and use $\alpha=0.1$ throughout.

To capture the essential dynamics of a SAF-SP on a tube, Eq.~(\ref{eq:ll-torque}) is projected onto a reduced set of collective modes. We work in the adiabatic approximation (AA), in which the dynamics is assumed to follow slow, soft modes, while all other faster modes relax instantaneously to their equilibrium configuration at fixed collective coordinates. Let $\vec{R}_{\ell}(t)=(R_{\ell x}(t),R_{\ell y}(t))$ be the center of the skyrmion in the layer $\ell$. For a current applied along $x$, the relevant slow variables are: $R_x(t)$, which represents the longitudinal position of the skyrmion pair, and the transverse separation $u_y(t)$ between the skyrmion centers in different layers. Due to the periodicity of the transverse direction, $u_y$ can be more conveniently represented by an angular coordinate $\theta$, as shown in Fig. \ref{fig:sketch-system}. Therefore, we define:
\begin{equation}
\label{eq:definition-of-slow-variables}
\begin{gathered}
R_x(t)=\frac{1}{2}\left[R_{1x}(t)+R_{2x}(t)\right],\\
u_y(t)=R_{1y}(t)-R_{2y}(t)\text{ (mod }N_y),\text{ }\theta(t)=\frac{2\pi}{N_y}u_y(t)\text{ (mod }2\pi), 
\end{gathered}
\end{equation}
where $N_y$ is the circumference of the tube. The separation is induced by the opposite SkHEs acting on the two skyrmions and, within the AA, is accompanied by a deformation of the spin textures. This is a reasonable assumption whenever $\alpha$ is not too small, the applied current is not too strong, and/or the domain-wall thickness is not too large -- cases in which additional internal modes can become dynamically relevant [SM]. Because the skyrmions deform along the adiabatic path, the centers extracted from the magnetization fields depend on the weighting function used to define them. Here, each layer-resolved center $\vec{R}_{\ell}$ is defined using the squared gradient of the magnetization as weights, $|\nabla\vec{m}_{\ell}|^2$, rather than the more commonly used topological charge density \cite{Kong2013}, since this is a choice that decouples the separation and shape deformation modes [SM].


Under these assumptions, the projected dynamics reduces to a set of two coupled collective-coordinate equations of motion [SM],
\begin{equation}
\label{eq:equations-of-motion}
\begin{gathered}
v_x(\theta)\equiv\dot{R}_x=\gamma H(\theta)+\zeta_1 j_x S(\theta),\text{ }\dot{\theta}=\alpha\gamma O(\theta)+\zeta_1 j_x Q(\theta)\\
\Rightarrow v_x(\theta)=\frac{S(\theta)}{Q(\theta)}\dot{\theta}+\gamma H(\theta)-\alpha\gamma\frac{O(\theta)S(\theta)}{Q(\theta)},
\end{gathered}
\end{equation}
where $v_x$ is the instantaneous pair drift velocity, and 
$H$, $O$, $S$, and $Q$ are nonlinear periodic  
functions obtained along a separation path in $\theta$ [SM]. In the AA, this path is constructed by Augmented Lagrangian minimization at prescribed SP separations, so that all spin degrees of freedom are relaxed at fixed $\theta$. For comparison, we also use a rigid approximation (RA), in which the single-layer skyrmion profiles are kept fixed and displaced rigidly relative to one another. The RA neglects shape relaxation within each layer but still changes the collective bilayer texture through the relative angular separation $\theta$, whereas the AA includes the $\theta$-dependent deformation induced by interlayer exchange.

The four periodic functions in Eq.~(\ref{eq:equations-of-motion}) separate the energy-driven and current-driven contributions to the dynamics. The terms $\gamma H(\theta)$ and $\alpha\gamma O(\theta)$ arise from the variation of the magnetic energy with $\theta$, caused primarily by the interlayer exchange. The function $H$ describes how this energy gradient generates longitudinal motion of the pair, whereas $O$ is a restoring term that governs the relaxation of the angular separation toward its stable values. The STT-induced response is given by $\zeta_1 j_x S(\theta)$ and $\zeta_1 j_x Q(\theta)$, where $S$ and $Q$ are the effective longitudinal and angular mobilities, respectively. By mirror symmetry, reversing the angular separation, $\theta\rightarrow-\theta$, reverses the direction of both the energy-driven drift and the restoring force, meaning $H$ and $O$ are odd functions. In contrast, the STT mobilities do not depend on the sign of the separation, so $S$ and $Q$ are even. Because the applied current also drives $\theta$, it continuously modulates all four functions; this makes the drift velocity $v_x$ in the SAF  a nonlinear function of the current. In the AA, this $\theta$-dependence also incorporates the deformation of the magnetization textures along the relaxed separation path.

\textit{Locked and winding regimes.---} In terms of the dimensionless time $\tau=t/T$, where $t$ is the real time and $T$ the period, the solutions to Eq.~(\ref{eq:equations-of-motion}) satisfy
\begin{equation}
\label{eq:winding-number-def}
\theta(\tau=1)=\theta(\tau=0)+2\pi k\text{,  }k\in\mathbb{Z}.
\end{equation}

With this boundary condition, the reduced dynamics exhibits two distinct motion regimes: the \textit{locked} regime, corresponding to $k=0$, and the \textit{winding} regime, characterized by $k\neq0$
\footnote{More formally, the angular separation takes values on the circle $S^1$. Upon defining $\tau=0$ and $\tau=1$, each cyclic trajectory defines a map $S^1\rightarrow S^1$ whose homotopy class is labeled by the integer $k\in\pi_1(S^1)=\mathbb{Z}$. This integer is the \textit{winding number}: $k=0$ identifies the trivial class, whereas the sign and magnitude of $k\neq0$ give the orientation and net number of revolutions around the tube.}. 

In the locked regime, the separation coordinate remains confined by the interlayer coupling, preventing the skyrmion pair from completing a revolution around the tube. Here, the angular separation $\theta$ is not necessarily fixed: a time-dependent current can also induce bounded motion satisfying $\theta(\tau=1)=\theta(\tau=0)$. However, over a complete locked cycle, the contribution associated with the angular motion -- i.e., the term $\propto{\theta}^{\prime}\equiv d\theta/d\tau$ of the velocity $v_x(\theta(\tau))$ in Eq. (\ref{eq:equations-of-motion}) -- averages to zero. Therefore, the average pair drift velocity $\bar{v}_x$ is given by
\begin{equation}
\label{eq:definition-of-guy}
\bar{v}_x=\int_0^1 v_xd\tau =\int_0^1\underbrace{\left[\gamma H(\theta(\tau))-\alpha\gamma\frac{O(\theta(\tau))S(\theta(\tau))}{Q(\theta(\tau))}\right]}_{\equiv\mathcal{G}(\theta(\tau))}d\tau.
\end{equation}

From Eq.~(\ref{eq:definition-of-guy}), it follows that, regardless of the current protocol, the locked regime is only possible up to a \textit{critical velocity} $\bar{v}_x^c$:
\begin{equation}
\label{eq:definition-of-critical-velocity}
\bar{v}_x^c=\max_{\theta\in[0,2\pi)}\mathcal{G}(\theta).
\end{equation}

Among all possible locked motions, a \textit{stationary} solution -- in which the pair separation is fixed, $\theta^{\prime}=0$ -- is obtained under a dc current drive $J_{dc}$. In this case, the separation $\theta=\theta^\star$ simultaneously satisfies $\mathcal{F}(\theta^\star)\equiv\alpha\gamma O(\theta^\star)+\zeta_1J_{dc}Q(\theta^\star)=0$ and $\bar{v}_x=\mathcal{G}(\theta^\star)$. If more than one root $\theta^\star$ exists, the stable one corresponds to $\mathcal{F}'(\theta^\star)<0$. In such a stationary locked motion, $J_{dc}$ is sufficiently small so that the SAF-SP does not overcome the point of maximal force, where $\partial^2E/\partial\theta^2=0$ for the total system energy $E$.


For average velocities exceeding the limit imposed by Eq.~(\ref{eq:definition-of-critical-velocity}), the applied current is sufficiently strong to drive the SAF-SP into a winding motion around the tube. Unlike the locked regime -- where stationary motion lacks an intrinsic period or the value of $T$ is set by a time-dependent current -- the period $T$ in the winding regime is naturally fixed by the time required for $\theta$ to complete a full rotation from $0$ to $2\pi$. For a given trajectory in this regime, $T$ can be computed as a function of $k$ and $\bar{v}_x$ as:
\begin{equation}
T[\theta]=\left(k\oint\frac{S(\vartheta)}{Q(\vartheta)}d\vartheta\right)\left[\bar{v}_x-\int_0^1\mathcal{G}(\theta(\tau))d\tau\right]^{-1},
\end{equation}
which results in the obvious limit of $T\rightarrow 0$ as $\bar{v}_x\rightarrow\infty$.   

\begin{figure*}[t!]
    \centering
    \includegraphics[width=1\linewidth]{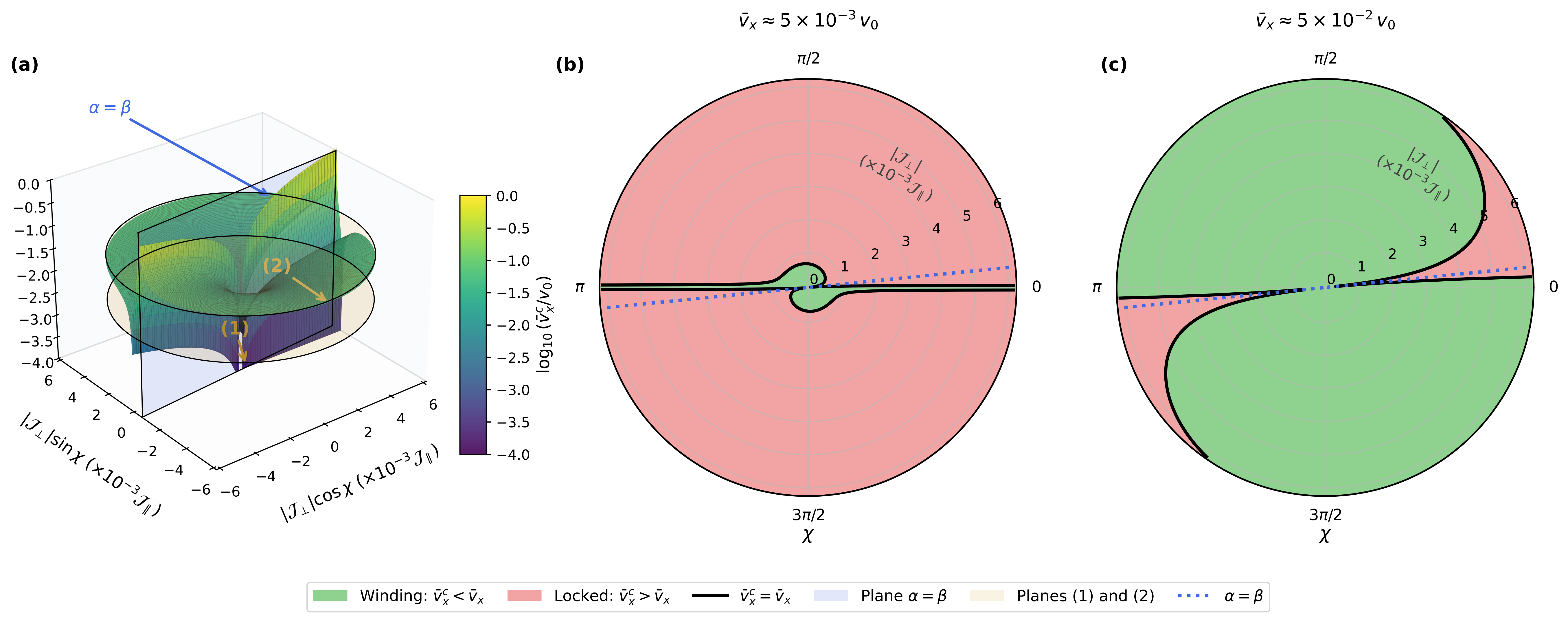}
    \caption{\textbf{Locked-to-winding boundary $\bar{v}_x^c$ as a function of the system/control parameters $|\mathcal{J}_{\perp}|$, $\chi$, and the average pair velocity $\bar{v}_x$.} (a) 3D representation of $\bar{v}_x^c$, with cutting planes \textit{(1)} and \textit{(2)} corresponding to selected target velocities $\bar{v}_x$, together with the plane $\alpha=\beta=0.1$, where the critical velocity diverges ($\bar{v}_x^c\rightarrow\infty$) and the system remains in the locked regime irrespective of $\bar{v}_x$; (b) 2D map of the locked (red) and winding (green) regions as a function of $\chi$, the polar coordinate, and $|\mathcal{J}_{\perp}|$, the radial coordinate, for $\bar{v}_x\approx5\times 10^{-3}v_0$, corresponding to plane \textit{(1)} in panel (a); (c) same as in (b), but for $\bar{v}_x\approx5\times 10^{-2}v_0$, corresponding to plane \textit{(2)} in panel (a). The black lines in (b) and (c) represent the computed $\bar{v}_x^c$ boundaries, while the dotted blue lines denote the $\beta=\alpha$ values. The characteristic velocity $v_0=\gamma\mathcal{J}_{\parallel}a/\mu\approx 4.14\times 10^3$ m$\cdot$s$^{-1}$ with the model parameters used, giving about $\bar{v}_x\approx20$ m$\cdot$s$^{-1}$ and $\bar{v}_x\approx200$ m$\cdot$s$^{-1}$ in panels (b) and (c), respectively.}
    \label{fig:vcrit-landscape}
\end{figure*}

Equations~(\ref{eq:definition-of-guy})-(\ref{eq:definition-of-critical-velocity}) determine whether a prescribed $\bar{v}_x$ can be achieved without winding; this criterion can be then used to map the two dynamical regimes in the parameter space. To systematically explore this space and clarify how the STT components $\zeta_1$ and $\zeta_2$ nontrivially delimit the two motion regimes, we introduce an angular variable $\chi$ to characterize their balance:
\begin{equation}
\label{eq:chi-definition}
\chi=\operatorname{atan2}(\zeta_2,\zeta_1),
\end{equation}
so that $\zeta_1=\zeta\cos\chi$ and $\zeta_2=\zeta\sin\chi$, with $\zeta=\sqrt{\zeta_1^2+\zeta_2^2}$ the total STT coupling magnitude. With this change of variables, it becomes possible to represent the structure of the locked-to-winding boundary $\bar{v}_x^c$ as a function of $|\mathcal{J}_{\perp}|$, $\chi$, and $\bar{v}_x$, as shown in Fig.~\ref{fig:vcrit-landscape}. The ridge at $\alpha=\beta$ corresponds to the singular limit in which the current does not excite the separation mode. This means that, in this case, the pair remains in the locked regime for any applied current and irrespective of $\bar{v}_x$, which is formally represented by $\bar{v}_x^c\rightarrow\infty$. In contrast, near $\beta=0$, $\bar{v}_x^c$ is strongly suppressed, allowing winding motion already at very low drift velocities -- a property that will be exploited below.

\textit{Optimal control problem.---} Having established which dynamical regimes are accessible by the SAF-SP, we are now interested in which current protocol realizes a prescribed $\bar{v}_x$ with the lowest Ohmic power. Apart from a common resistance-dependent prefactor, the power is defined as
\begin{equation}
\label{eq:power-def}
\mathcal{P}=\int_0^1 j_x^2(\tau)d\tau.
\end{equation}

For given $\bar{v}_x$ and $k$, the minimization of Eq. (\ref{eq:power-def}) is carried out on current profiles and angular trajectories that satisfy Eq. (\ref{eq:equations-of-motion}) and the corresponding cyclic boundary condition given by Eq. (\ref{eq:winding-number-def}). This problem can be more conveniently formulated in terms of $\theta(\tau)$, instead of the control current $j_x$. Solving Eq. (\ref{eq:equations-of-motion}) for $j_x$ gives 
\begin{equation}
\label{eq:jx-from-eq-motion}
j_x(\tau)=\frac{1}{T\zeta_1}p(\tau)+h(\tau),
\end{equation}
where $p(\tau)\equiv\theta'(\tau)[Q(\theta(\tau))]^{-1}$, and   $h(\tau)\equiv-(\alpha\gamma\zeta_1^{-1})O(\theta(\tau))[Q(\theta(\tau))]^{-1}$. The first term accounts for the angular motion, with $p$ acting as a control-weighted winding rate, whereas $h$ 
is the current required to keep the separation fixed at a given value of $\theta$. 

In the locked regime, it can be shown that among all possible protocols, a dc current $J_{dc}$, corresponding to the stationary solution, minimizes the Ohmic power losses \footnote{For $k=0$, the average pair drift velocity is $\bar{v}_x=\int_0^1\mathcal{G}(\theta(\tau))d\tau$. Since 
\begin{equation}
\int_0^1\left[\frac{\mathcal{G}(\theta(\tau))}{Q(\theta(\tau))}\right]\theta'(\tau)d\tau=0
\end{equation}
along a closed trajectory, from Eq. (\ref{eq:power-def}) one can write
\begin{equation}
\mathcal{P}=\left(\frac{\alpha}{\zeta_2}\right)^2
\int_0^1\mathcal{G}^2(\theta(\tau))d\tau+\frac{1}{T^2\zeta_1^2}\int_0^1\frac{\theta'^2(\tau)}{Q^2(\theta(\tau))}d\tau\geq
\left(\frac{\alpha\bar{v}_x}{\zeta_2}\right)^2,
\end{equation}
where the inequality follows from Cauchy-Schwarz. The equality requires $\theta'(\tau)=0$, which corresponds to the stationary solution and a constant current $J_{dc}=\alpha\bar{v}_x/\zeta_2=\alpha\bar{v}_x/(\zeta\sin\chi)$.}.  Similarly, we can derive the general character of the optimal protocols in the winding regime. In this case, as $\bar{v}_x\rightarrow\infty$, the $p$ term on the right-hand side of Eq.~(\ref{eq:jx-from-eq-motion}) -- proportional to $T^{-1}$ -- clearly dominates over $h$. To leading order, minimizing the power gives $\mathcal{P}_{\mathrm{opt}}=\min_{p(\tau)}\int_0^1j_x^2d\tau\approx(\mathcal{C}_k/T\zeta_1)^2$, where $\mathcal{C}_k$ is a winding constant \footnote{Because $Q$ is $2\pi$-periodic and nonzero along a winding trajectory,
\begin{equation}
\label{eq:winding-constant-definition}
\int_0^1p(\tau)d\tau = \int_{\theta(0)}^{\theta(0)+2\pi k}\frac{d\vartheta}{Q(\vartheta)}\equiv\mathcal{C}_k,
\end{equation}
where $\mathcal{C}_k$ is fixed by the winding number $k$. Therefore, to leading order,
\begin{equation}
\mathcal{P}
\approx
\frac{1}{T^2\zeta_1^2}\int_0^1p^2(\tau)d\tau
\geq
\frac{1}{T^2\zeta_1^2}
\left(\int_0^1p(\tau)d\tau\right)^2
=
\left(\frac{\mathcal{C}_k}{T\zeta_1}\right)^2,
\end{equation}
using Cauchy-Schwarz, with the equality holding for $p(\tau)=\mathcal{C}_k$.}. Thus, the optimal current $j_x^{\mathrm{opt}}\approx\mathcal{C}_k/(T\zeta_1)$ also tends to a dc limit as $\bar{v}_x\rightarrow\infty$. Therefore, genuinely time-dependent optimal protocols emerge at \textit{intermediate} average drift velocities: above the locked regime, but below the high-velocity winding limit where the optimum becomes effectively dc.

\textit{Benchmark against dc driving.---} In experiments, it is often the case where SAF-SPs are moved via dc current pulses \cite{Dohi2019,Pham2024}; therefore, quantitatively accessing the power consumption advantage of the optimal control protocol (OCP) over dc currents is an experimentally relevant question.  For $\bar{v}_x<\bar{v}_x^c$ within the locked regime, the optimal solution of Eq. (\ref{eq:equations-of-motion}) already requires a constant current, where skyrmions are separated by a stable angular separation $\theta^{\star}$ satisfying $\bar{v}_x=\mathcal{G}(\theta^{\star})$. Thus, in this case, the power ratio between the dc and optimal protocols, by construction, is $\mathcal{P}_{dc}/\mathcal{P}_{\rm opt}=J_{dc}^2/\mathcal{P}_{\rm opt}^{\rm locked}=1$.

However, for $\bar{v}_x>\bar{v}_x^c$, unlike in the locked regime, the optimal protocol in the winding regime can be time-dependent, becoming effectively dc only in the high-velocity limit. To quantify its advantage over the dc drive commonly used in experiments, we now consider, as a reference, a winding trajectory driven by a constant current that produces the same $\bar{v}_x$. Setting $j_x(\tau)\equiv j_{dc}$ in Eq. (\ref{eq:equations-of-motion}) gives $\mathcal{F}(\theta)=(1/T)\theta^{\prime}$, so that the dc trajectory has a fixed point $\theta^{\prime}=0$ whenever $j_{dc}$ intersects the function $h$. For all possible angular separations $\theta$, we can therefore define
\begin{equation}
\label{eq:depinning-current}
J_{\rm dep}=\max_{\theta\in[0,2\pi)}|h(\theta)|=\max_{\theta\in[0,2\pi)}\left|\frac{\alpha\gamma}{\zeta_1}\frac{O(\theta)}{Q(\theta)}\right|
\end{equation}
as the \textit{effective depinning current}, in which $|j_{dc}|>J_{\rm dep}$ allows for the SAF-SP to explore all possible separations  and complete a lap under constant current, without being trapped. In fact, in the limit of $j_{dc}\rightarrow J_{\rm dep}$, the total travel time diverges, and the SAF-SP spends almost the entire cycle near the bottleneck angle $\theta^b$, which  solves $|h(\theta^b)|=J_{\rm dep}$ -- or, in other words, near the angle that maximizes the function $h$. 

As a consequence, one way to find regions of interest in the parameter space where choosing the OCP is advantageous over dc currents involves, not exclusively, obtaining conditions in which the critical velocity $\bar{v}_x^c$ is minimized to guaranty a wider range of possible velocities $\bar{v}_x$ that hit the winding regime, while simultaneously keeping a large depinning dc current limit $J_{\rm dep}$. In this sense, the limit of $\beta\rightarrow0$ and large $|\mathcal{J}_{\perp}|$ does exactly that. 

We can understand this as follows. When $\beta=0$, the function $\mathcal{G}$ vanishes exactly due to identities of the periodic functions in Eq. (\ref{eq:equations-of-motion}) [see SM and Fig. (\ref{fig:vcrit-landscape})]. From Eq. (\ref{eq:definition-of-critical-velocity}), it follows that $\bar{v}_x^c=0$ or, in other words, in the $\beta=0$ limit the skyrmion pair \textit{must} undergo winding for any nonzero prescribed $\bar{v}_x$. The gyrotropic forces from the two layers cancel when in a compensated locked regime, and the longitudinal drift driven by the STT is controlled by the nonadiabatic component. Thus, when $\beta=0$, a stationary locked solution cannot sustain longitudinal motion; a nonzero drift requires a cycle of the separation coordinate, which may be driven either by a dc current above the depinning threshold or by a time-dependent protocol. For larger $|\mathcal{J}_{\perp}|$ values, the $\bar{v}_x^c=0$ limit when $\beta=0$ remains unaffected, but $J_{\rm dep}$ is increased: one can show that $J_{\rm dep}\sim V_x/M_{xx}$ when $\beta\rightarrow0$ and $\alpha$ is finite, where $V_x$ and $M_{xx}$ are the magnetization field integrals defined in [SM]. Since $V_x$ depends on the internal field $\vec{B}_{\ell}$, the numerator increases with $|\mathcal{J}_{\perp}|$ -- and so $J_{\rm dep}$. 

\begin{figure*}[t!]
    \centering
    \includegraphics[width=1\linewidth]{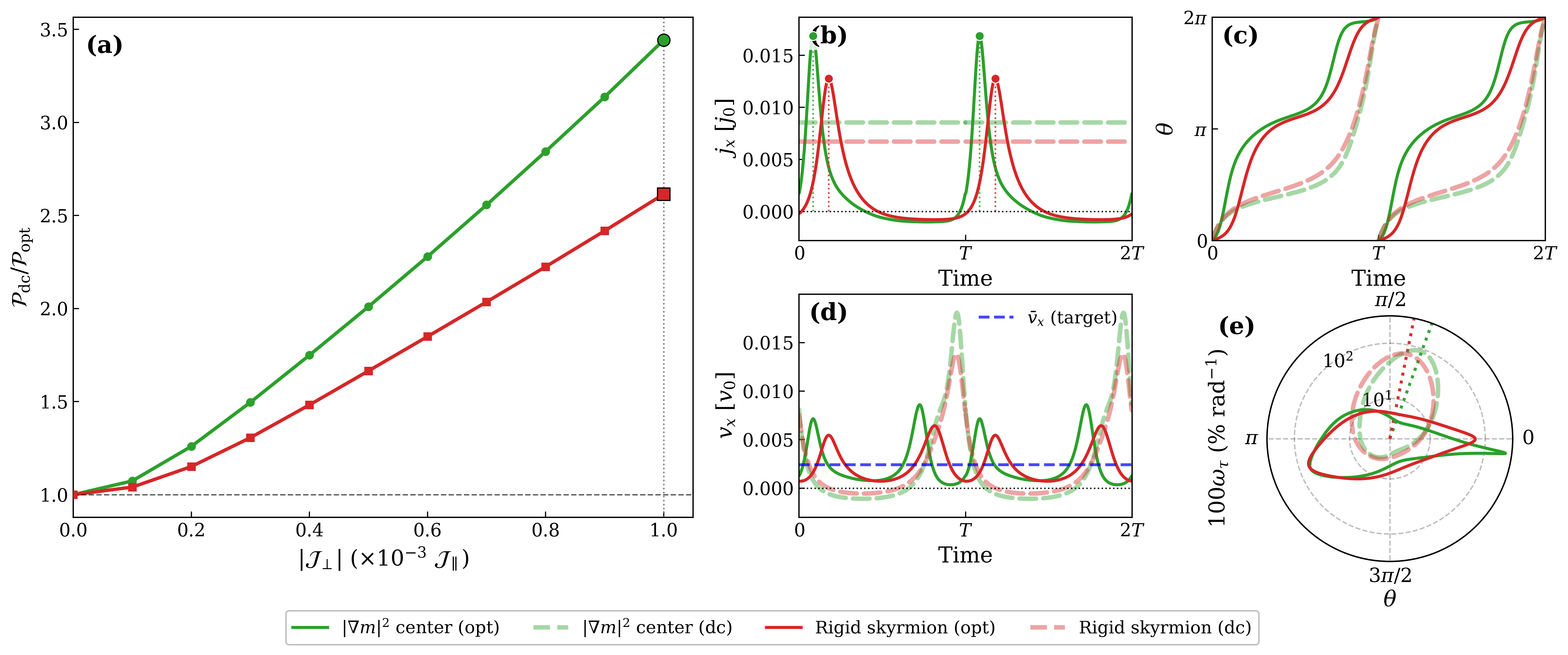}
    \caption{\textbf{dc and optimal control protocols computed for an average drift velocity $\bar{v}_x\approx2.5\times 10^{-3}v_0$, $\beta\approx-0.022$, and $\alpha=0.1$, within the rigid and adiabatic approximations.} (a) ratio $\mathcal{P}_{\rm dc}/\mathcal{P}_{\rm opt}$ as a function of the interlayer exchange magnitude, with the rigid and adiabatic results shown in red and green, respectively. (b-d) comparison between the optimal control and dc protocols for the highlighted end points in (a), showing the instantaneous applied current, the separation coordinate $\theta$, and the instantaneous drift velocity $v_x$ of the skyrmion pair over two full periods $2T$. (e) temporal weight profile $\omega_{\tau}(\theta)\equiv d\tau/d\theta$, which quantifies the fraction of the period allocated per unit angular separation coordinate $\theta$, such that $\int_0^{2\pi}\omega_{\tau}(\theta)d\theta=1$. In (b-e), green/red solid and long-dashed curves denote the optimal and dc protocols, respectively. In panel (b), the colored dots mark the instants at which the optimal trajectories cross their respective bottleneck angles $\theta^b$, which maximize the pinning current function $h$ (see text), also represented by the dotted lines in panel (e). In (d), the blue dotted line indicates the average velocity $\bar{v}_x$. Finally, $v_0=\gamma\mathcal{J}_{\parallel}a/\mu\approx 4.14\times 10^3$ m$\cdot$s$^{-1}$ and $j_0=v_0/\zeta\approx 6.6\times10^{14}$ A/m$^2$ are the characteristic velocity and current, in this order.}
    \label{fig:opt-and-dc}
\end{figure*}

In this case, the advantage of the OCP to the dc protocol mainly arises because, while the dc protocol is bounded from below by the current needed to overcome the fixed point, a time-dependent optimal protocol can cross the bottleneck angle only within a small fraction of the cycle, and distribute the remaining dynamics in lower-cost regions, leading to a smaller average power. This characterizes a clear example where the separation mode is useful and effectively employed to reduce the heating power w.r.t. an experimentally relevant protocol. Fig. \ref{fig:opt-and-dc} shows the ratio $\mathcal{P}_{dc}/\mathcal{P}_{\rm opt}$ as a function of $|\mathcal{J}_{\perp}|$ for a given $\beta$ in the relatively small-$\beta$ regime ($|\beta|<\alpha$), for both RA and AA, as well as the comparison between the properties of each trajectory. In particular, Fig. \ref{fig:opt-and-dc} (b) illustrates the mechanism behind this advantage: the OCP concentrates its higher current magnitudes near the fixed point, maintaining low-current dynamics elsewhere.

As can be seen from the numerical calculations, close to the $\beta=0$ condition, the ratio $\mathcal{P}_{dc}/\mathcal{P}_{\rm opt}$ strongly increases as a function of the interlayer exchange magnitude, reaching significant values and, as expected, approaching unity as $|\mathcal{J}_{\perp}|\rightarrow0$. The comparison between RA (red) and AA (green) with the same model parameters, namely $\alpha$, $\beta$, $|\mathcal{J}_{\perp}|$, and $\bar{v}_x$, also shows an interesting result: the role played by the skyrmion deformation induced by interlayer exchange to reach even higher power ratios. Since in the AA skyrmions experience a size reduction to minimize the interlayer exchange energy, the profile of the pinning current function $h$ becomes more nonuniform than in RA, i.e., exhibiting narrower peaks separated by extended low-valued regions. Pointwise depinning can be achieved whenever $j_x>h$, and the optimal control solution advantage is obtained by dynamically decreasing $j_x(\tau)$ in such low-valued regions. This rationale can be extended to racetracks with larger widths $N_y$, in which such low-valued regions of $h$ are enlarged and the advantage over the dc current is expected to be even greater.

\textit{Minimal power.---} Next, we analyze the minimal power required to reach a \textit{fixed} average velocity $\bar v_x$, with $\mathcal P_{\rm opt}$ evaluated as a function of $|\mathcal J_{\perp}|$ and of $\chi$. This choice is motivated by the tunability of real SAF systems: while $\beta$ is mostly set by material dependent spin-transport properties, the interlayer exchange can be varied more directly by changing the spacer thickness through the RKKY coupling between the ferromagnetic layers \cite{Legrand2019}, or even at optical means \cite{Ma2025}. It is therefore natural to ask how the optimal power changes with $|\mathcal J_{\perp}|$,  $\mathcal P_{\rm opt}(|\mathcal J_{\perp}|)$, and whether the minimum occurs in the locked or winding regime.

For given $\chi$ and $\bar v_x$, the locked sector provides a simple reference. Since the minimum-power solution within this sector is stationary,  with fixed $\theta$, and using the periodic function identities [SM] and Eq. (\ref{eq:definition-of-guy}), one obtains
\begin{equation}
\label{eq:power-opt-locked}
\mathcal{P}_{\rm opt}^{\rm locked}=J_{dc}^2=\left(\frac{\alpha\bar{v}_x}{\zeta\sin\chi}\right)^2,
\end{equation}
and therefore $\mathcal{P}_{\rm opt}^{\rm locked}$ is independent of $|\mathcal{J}_{\perp}|$ \footnote{The dependence of $\mathcal{P}_{\rm opt}^{\rm locked}$ on $\beta$ (or $\chi$) is simultaneously implicit and explicit, since in the Zhang-Li STT model \cite{Zhang2004} $\zeta_1$ is itself a function of $\beta$: $\zeta_1\equiv\zeta_1(\beta)\propto(1+\beta^2)^{-1}$}. In other words, within the dynamical model given by Eq. (\ref{eq:equations-of-motion}), the interlayer exchange controls the existence and position of the stable stationary locked solution $\theta^{\star}$, but not its power once such a solution exists. Therefore, this regime defines a dc baseline in the $\mathcal P_{\rm opt}(|\mathcal J_{\perp}|)$ landscape.

The winding regime can then be directly compared with this baseline. When $\mathcal P_{\rm opt}^{\rm winding}<\mathcal P_{\rm opt}^{\rm locked}$, the separation mode is not merely dynamically allowed, but provides a lower power route than the best dc solution at the same $\chi$ and $\bar v_x$. Two distinct mechanisms can lead to this situation. The first was previously discussed and occurs in the limit of $\chi\to0$. In this case, Eq.~(\ref{eq:power-opt-locked}) shows that the $\mathcal{P}_{\rm opt}^{\rm locked}$ baseline becomes singular, while the locked-to-winding boundary is strongly suppressed, $\bar{v}_x^c\to 0$. Thus, the longitudinal transport of the skyrmion pair is more efficiently sustained by cycling of the separation coordinate, naturally leading to $\mathcal P_{\rm opt}^{\rm winding}<\mathcal P_{\rm opt}^{\rm locked}$. 

\begin{figure*}[t]
    \centering  \includegraphics[width=1\linewidth]{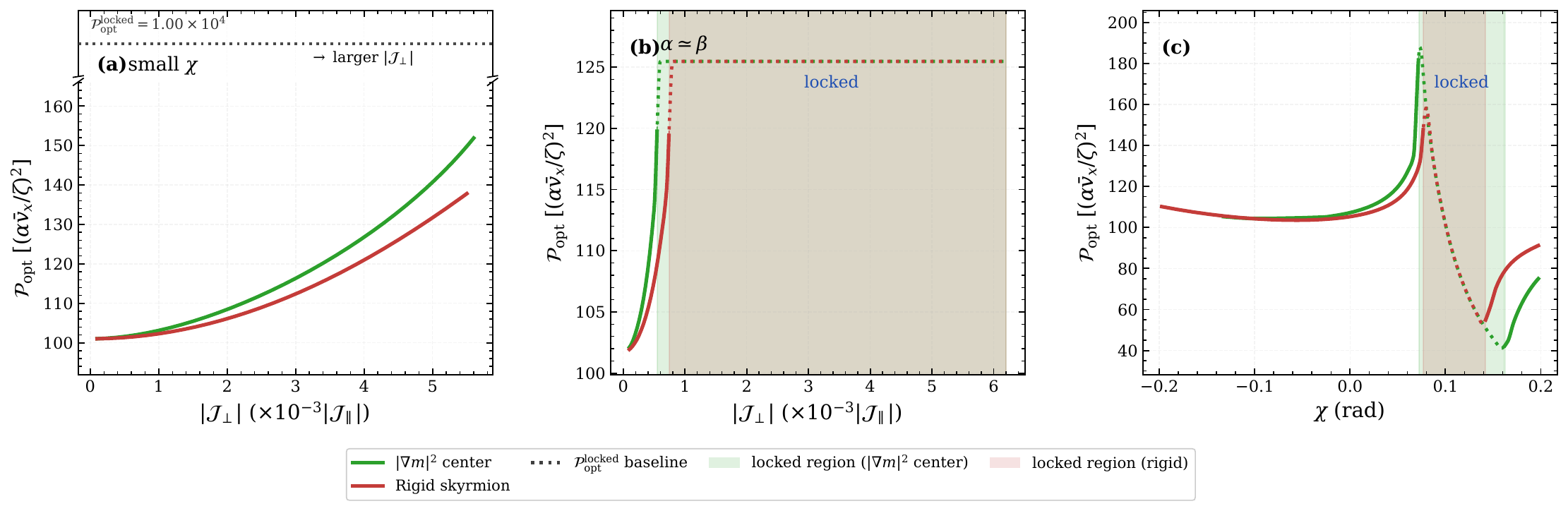}
    \caption{\textbf{Optimal protocol power for an average drift velocity $\bar{v}_x\approx 5\times10^{-2}v_0$  and $\alpha=0.1$, in the rigid and adiabatic approximations.} (a) $\mathcal{P}_{\rm opt}$ as a function of $|\mathcal{J}_{\perp}|$ in the small $\chi$ regime, with $\chi\approx0.01$ rad. (b) $\mathcal{P}_{\rm opt}$ as a function of $|\mathcal{J}_{\perp}|$ near $\alpha\simeq\beta$, here for $\beta=0.09$. (c) $\mathcal{P}_{\rm opt}$ as a function of $\chi$ for $|\mathcal{J}_{\perp}|=2\times 10^{-3}\mathcal{J}_{\parallel}$. Dotted lines denote the $\mathcal{P}_{\rm opt}^{\rm locked}$ baseline (see text), while shaded regions mark the corresponding locked regimes. In (a), $\mathcal{P}_{\rm opt}^{\rm locked}$ lies about two orders of magnitude above the winding branch and is reached only at larger $|\mathcal{J}_{\perp}|$ values outside the plotted range, as indicated by the broken vertical axis.}
    \label{fig:minimal-power}
\end{figure*}

A second mechanism, qualitatively different, appears near $\beta\approx\alpha$, but only for sufficiently weak $|\mathcal J_{\perp}|$. Although the factor $Q(\theta)$ suppresses the current-induced separation dynamics close to $\beta=\alpha$, reducing $|\mathcal J_{\perp}|$ softens the separation mode and can bring the winding sector back into reach. Unlike the $\chi\to0$ case, the gain here does not come from the divergence of $\mathcal P_{\rm opt}^{\rm locked}$, but from the reduced energy cost of cycling a soft internal coordinate. Figures \ref{fig:minimal-power} (a,b) illustrate these cases: in (a), the small-$\chi$ limit ($\chi\approx0.01$ rad), and in (b) a case in which $\beta\approx\alpha$. As expected, both the optimal powers of RA and AA coincide when $|\mathcal{J}_{\perp}|=0$.

Apart from the experimentally motivated analysis of the $\mathcal P_{\rm opt}(|\mathcal J_{\perp}|)$ landscape, it is also useful to understand where the lowest power occurs in the full parameter space for a given $\bar v_x$, but free $\chi$ and $|\mathcal J_{\perp}|$. From Eq.~(\ref{eq:power-opt-locked}), $\mathcal P_{\rm opt}^{\rm locked}\propto\sin^{-2}\chi$, and therefore the locked power decreases monotonically with increasing $|\sin\chi|$. Thus, within each locked regime sector, the lowest power value is reached at the boundary with the largest $|\sin\chi|$, while the opposite boundary carries the largest value, as can be seen in the shaded region of Fig. \ref{fig:minimal-power} (c). Such boundaries are shared with the neighboring winding sectors: close to the locked-to-winding transition, the winding period diverges and the trajectory spends most of the cycle near the separation $\theta^\star$, so that $\mathcal P_{\rm opt}^{\rm winding}$ tends to $\mathcal P_{\rm opt}^{\rm locked}$. This explains why the extrema of the power landscape in Fig.~\ref{fig:minimal-power} (c) accumulate near the locked-to-winding boundaries. 

Therefore, the minimum of the full locked sector is obtained for a purely nonadiabatic torque, $\chi=\pi/2$ rad, and defines the natural power scale
$\mathcal P^\star=
(\alpha\bar v_x/\zeta)^2$ for the whole parameter space. An important consequence is that this low power regime does not correspond to the trivial limit of uncoupled skyrmions, $|\mathcal J_{\perp}|=0$, being smaller by at least a factor $\alpha^2$. Indeed, any locked solution satisfying $\sin\chi>\alpha$ is already more efficient than the case of uncoupled skyrmions \footnote{In the uncoupled limit, the longitudinal equation reduces to $\bar v_x=\zeta_1 j_x S_0$, where
\begin{equation}
S_0=S(\theta=0)=\frac{1+\alpha\beta\rho}{1+\alpha^2\rho}\text{,  }
G_{XY}^2=\rho M_{XX}M_{YY}\text{,  }
0\leq\rho\leq1.
\end{equation}
The bound on $\rho$ follows from Cauchy-Schwarz [SM]. Using $\zeta_1=\zeta\cos\chi$ and $\beta=\tan\chi$, the uncoupled power becomes
\begin{equation}
\mathcal P_{\rm unc}(\chi)=
\left[
\frac{(1+\alpha^2\rho)\bar v_x}
{\zeta(\cos\chi+\alpha\rho\sin\chi)}
\right]^2.
\end{equation}
Minimization over $\chi$ gives
\begin{equation}
\min_{\chi}\mathcal P_{\rm unc}=
\left[
\frac{(1+\alpha^2\rho)\bar v_x}
{\zeta\sqrt{1+\alpha^2\rho^2}}
\right]^2,
\end{equation}
and therefore
\begin{equation}
\left(\frac{\bar v_x}{\zeta}\right)^2
\leq
\min_{\chi}\mathcal P_{\rm unc}
\leq
\left(\frac{\sqrt{1+\alpha^2}\bar v_x}{\zeta}\right)^2.
\end{equation}
On the other hand, Eq.~(\ref{eq:power-opt-locked}) implies $\mathcal P_{\rm opt}^{\rm locked}<(\bar v_x/\zeta)^2$ whenever $\sin\chi>\alpha$. Since $\mathcal P_{\rm opt}^{\rm winding}\rightarrow\mathcal P_{\rm opt}^{\rm locked}$ upon approaching a locked-to-winding boundary from the winding side, the same strict inequality holds in a finite neighborhood of that boundary within the winding sector.} -- and, consequently, whenever such a locked solution borders a winding sector, there are winding protocols whose power is also lower than that of uncoupled skyrmions. Therefore, a finite $|\mathcal{J}_{\perp}|$ is required to create the coupled internal coordinate that allows the SAF-SP to approach the minimal power regime.

Taken together, these results separate three notions of optimality. The uncoupled limit, $|\mathcal J_{\perp}|=0$, provides the isolated skyrmion reference. The locked sector sets the minimal power for the skyrmion pair transport in the whole parameter space, $\mathcal P^\star$. Finally, the winding sector can be \textit{locally optimal}, outperforming the best locked/dc solution available at same $\chi$; therefore, the winding sector exhibits finite, physically accessible regions where optimal control turns the internal separation mode of SAF-SPs into a functional low-dissipation transport channel.






\textit{Conclusions and outlook.---} In conclusion, we have solved the optimal current problem for transporting skyrmion pairs in SAFs within a reduced two-coordinate model, minimizing the Ohmic losses for a prescribed target velocity. The tube geometry plays a central role: it makes the transverse separation a periodic internal coordinate, so that the pair motion is not a rigid drift, but results from the coupled evolution of translation and separation. This coupling makes the velocity a nonlinear functional of the applied current, which is precisely what allows the transport to be optimized. The reduced dynamics contains two distinct regimes: locked and winding. The locked branch sets the lowest power scale for the pair transport, whereas the winding branch can be locally more efficient, outperforming the locked/dc solution available at the same STT balance $\chi$. The analysis also shows that finite interlayer exchange is essential: by creating a coupled internal coordinate, SAF skyrmion pairs can reach lower powers than isolated skyrmions. Thus, optimal control exposes a hidden dynamical resource: the SkHE, usually regarded as detrimental, can be converted into transport with reduced dissipation by efficiently using the internal separation mode. These results provide guiding principles and identify a qualitatively distinct setting for skyrmion media with lower power losses, showing an example of how internal modes of topological spin textures can be exploited -- rather than suppressed -- in transport protocols.

\section*{Acknowledgements}
The authors would like to thank T. Sigurj\'onsd\'ottir for helpful discussions and useful comments. 
This work was supported by the Icelandic Research Fund (Grants No. 2410333 and No. 217750), the University of Iceland Research Fund (Grant No. 15673), the Faculty of Technology and the Department of Mathematics and Physics at Linnaeus University (Sweden), the Swedish Research Council (Grant No. 2020-05110), the Russian Science Foundation (Grant no. 23-72-10028), and the Crafoord Foundation (Grant No. 20231063). The computational resources were provided by the National Academic Infrastructure for Supercomputing in Sweden (NAISS), funded by the Swedish Research Council.
\vspace{-5mm}

\bibliographystyle{apsrev4-2}
\bibliography{bib}





\end{document}